\documentclass[aps,prl,amsmath,amssymb,twocolumn,superscriptaddress,showpacs,showkeys]{revtex4-1}

\usepackage{latexsym}
\usepackage{bm}
\usepackage{amssymb}
\usepackage{amsmath}
\usepackage{graphicx}% Include figure files
\usepackage{dcolumn}% Align table columns on decimal point
\usepackage{bm}% bold math
\usepackage{gensymb}
\usepackage{braket}
\usepackage[dvipsnames]{xcolor}
\usepackage[normalem]{ulem}

\begin{document}

%\preprint{}


%\title{A high sensitivity trapped atom interferometer}
\title{Squeezing on momentum states for atom interferometry}



\author{Leonardo Salvi}
\affiliation{Dipartimento di Fisica e Astronomia and LENS -
Universit\`{a} di Firenze, INFN - Sezione di Firenze, Via Sansone 1, 50019 Sesto Fiorentino,~Italy}
\author{Nicola Poli}
\affiliation{Dipartimento di Fisica e Astronomia and LENS -
Universit\`{a} di Firenze, INFN - Sezione di Firenze, Via Sansone 1, 50019 Sesto Fiorentino,~Italy}
\author{Vladan Vuleti\ifmmode \acute{c}\else \'{c}\fi{}}
\email{vuletic@mit.edu}
\affiliation{Department of Physics, Research Laboratory of Electronics, Massachusetts Institute of Technology, Cambridge, Massachusetts 02139, USA}
\author{Guglielmo M. Tino}
\email{guglielmo.tino@unifi.it}
\affiliation{Dipartimento di Fisica e Astronomia and LENS -
Universit\`{a} di Firenze, INFN - Sezione di Firenze, Via Sansone 1, 50019 Sesto Fiorentino,~Italy}

\date{\today}

\begin{abstract}
We propose and analyse a method that allows for the production of squeezed states of the atomic center-of-mass motion that can be injected into an atom interferometer. Our scheme employs dispersive probing in a ring resonator on a narrow transition of strontium atoms in order to provide a collective measurement of the relative population of two momentum states. We show that this method is applicable to a Bragg diffraction-based atom interferometer with large diffraction orders. The applicability of this technique can be extended also to small diffraction orders and large atom numbers by inducing atomic transparency at the frequency of the probe field, reaching an interferometer phase resolution scaling $\Delta\phi\sim N^{-3/4}$, where $N$ is the atom number. We show that for realistic parameters it is possible to obtain a 20 dB gain in interferometer phase estimation compared to the Standard Quantum Limit.
\end{abstract}

\maketitle

%%\section{Introduction}
A major effort in the field of atom interferometry \cite{Varenna2013} is focused on increasing the instrument sensitivity, either by enhancing the momentum transferred by the light onto the atoms \cite{PhysRevLett.100.180405,Kovachy2015} or by increasing the interrogation time \cite{Kovachy2015,PhysRevA.94.043608,PhysRevA.85.013639,PhysRevA.88.031605}. By applying differential schemes, many systematics and noise sources can be efficiently rejected as common-mode effects \cite{PhysRevA.89.023607,Rosi2014}, and one eventually meets the fundamental atom shot noise limit that arises from the uncorrelated phase noise of different atoms. The minimum phase resolution allowed by uncorrelated atomic states is the Standard Quantum Limit (SQL) $\Delta\phi_{\mathrm{SQL}}=1/\sqrt{N}$, where $N$ is the atom number. This limit can be overcome by introducing correlations between the individual particles, thereby producing squeezed atomic states, potentially reaching the Heisenberg limit $\Delta\phi_\mathrm{H}=1/N$ \cite{Wineland1992,PhysRevA.50.67}. Many schemes have been studied both theoretically \cite{PhysRevA.77.063811,PhysRevLett.94.023003} and experimentally \cite{ce8e7a60162111df803f000ea68e967b,PhysRevLett.104.073602,PhysRevLett.104.073604,McConnell2015,G.2014,Strobel424,PhysRevLett.113.103004,Riedel2010,Colangelo2017} with about 20 dB noise reduction compared to the SQL \cite{Hosten2016,PhysRevLett.116.093602}. The key feature of most of these schemes is the enhanced atom-light interaction in an optical resonator. Many experiments with optical resonators involve trapping the atoms in an optical potential generated by the resonator itself and inducing atomic correlations through internal atomic states. Because of these features, optical resonators are well suited for atomic clocks beyond the SQL. 
The implementation of these methods in atom interferometry remains, however, a challenging task.

In this Letter we propose and analyse a scheme that generates squeezed momentum states \cite{Wineland1992,Ma201189} for atom interferometry. In particular, we consider the production of squeezed states of the atomic center-of-mass motion by dispersive probing of a momentum-state superposition of ultracold strontium (Sr) atoms in an optical ring resonator. 
On the one hand, for the bosonic isotopes of strontium, the choice of the atomic species is motivated by its expected immunity to stray fields in atom interferometers and by the possibility of attaining long coherence times in quantum interference \cite{PhysRevLett.106.038501,PhysRevLett.113.023005}. On the other hand, the presence of narrow intercombination transitions makes the atom well suited for squeezing experiments involving external degrees of freedom.
\begin{figure}[htbp]\begin{center}
\includegraphics[width=0.5\textwidth]{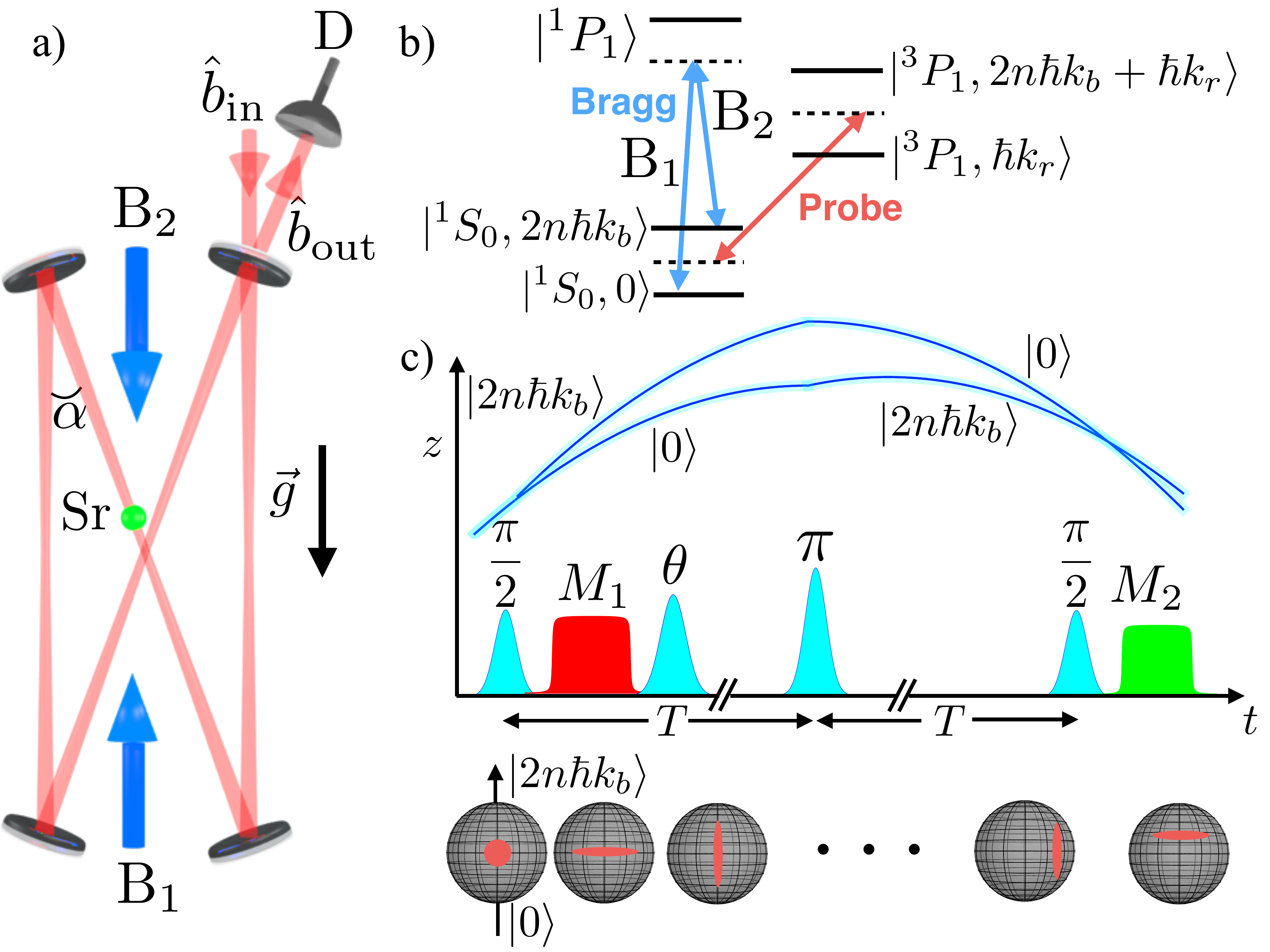}
\caption{a) Experimental setup with the probe field (red beams) coupled to a single mode of an optical ring cavity and interacting with the atomic system (green circle). The angle between gravity and the oblique laser fields is indicated as $\alpha$. The annihilation operators for the cavity input and ouput fields are indicated as $\hat{b}_{\mathrm{in}}$ and $\hat{b}_{\mathrm{out}}$, respectively. The photon phase shift is measured by monitoring the cavity reflected field through the detector D. The two Bragg beams ($\mathrm{B}_1$ and $\mathrm{B}_2$) propagate in free space and induce a coherent superposition of momentum states. b) Level diagram for probing on the intercombination transition $^1S_0-$$^3P_1$ (red arrow) and for momentum state manipulation through Bragg interaction on the $^1S_0-$$^1P_1$ transition (blue arrows). c) Interferometer trajectories and measurement sequence. The blue $\pi$ and $\pi/2$ pulses, separated by the interferometer time $T$, act as mirrors and beam splitters in a Mach-Zehnder interferometer, while the momentum state populations are measured through the probe beam in $M_1$. A Bragg light pulse, indicated as $\theta$, is used to generate a phase-sensitive state. The final populations of the interferometer are measured in $M_2$. In the lower part, the state evolution on the Bloch sphere corresponding to the two momentum states is illustrated. }
\label{fig.figure1}
\end{center}
\end{figure}
%%We consider the situation depicted in Fig.~\ref{fig.figure1} a) where a probe laser field at frequency $\omega$ is tuned close to the resonance frequency $\omega_0$ of the transition from ground $\ket{g}$ to excited $\ket{e}$ state of $N$ two-level atoms, with detuning $\Delta = \omega-\omega_0$. The probe laser is coupled to a single mode of an optical ring cavity as shown in Fig.~\ref{fig.figure1} b). This geometry allows for a more homogeneous coupling of the atoms with the probe field compared to linear cavities because of the absence of standing waves. Moreover, this setup allows the atoms to exit the cavity volume for atom interferometers in free space. The cavity resonance frequency is denoted as $\omega_c$ and the corresponding detuning of the probe field is $\delta=\omega-\omega_c$.

The proposed scheme is illustrated in Fig.~\ref{fig.figure1}, where two vertical counterpropagating laser fields $\mathrm{B}_1$ and $\mathrm{B}_2$ in free space induce the momentum state superposition between the states $\ket{^1S_0,p=0}$ and $\ket{^1S_0,p=2n\hbar k_b}$ by $n$-th order Bragg diffraction \cite{PhysRevLett.100.180405} on the dipole-allowed $^1S_0$ $-$ $^1P_1$ blue Sr transition at 461 nm. Here the atomic linear momentum is indicated by $p$ and the photon momentum is denoted by $\hbar k_b$.
We consider squeezing of the atomic states by collective measurements of the relative population of the two momentum states through dispersive detection in a ring cavity (Fig.~\ref{fig.figure1}). This is achieved by probing on the red $^1S_0-$$^3P_1$ intercombination line of strontium at 689 nm using a laser beam, of angular frequency  $\omega_r$, incident onto the cavity. Probing is performed when the free-falling atoms cross the cavity mode volume. 
The small linewidth $\Gamma=2\pi\times 7.6$ kHz of the intercombination transition allows one to distinguish the two momentum states via the Doppler effect which leads to a splitting of the resonance by $2\delta\omega_r = 2\pi n\times 28.6$ kHz.

In the following, measurements of the cavity output field $\hat{b}_{\mathrm{out}}$ are considered and the sensitivity to atom number fluctuations between momentum states is computed. As such measurements provide collective information about the ensemble without distinguishing between individual atoms, they project the ensemble into a collective state which corresponds to the measurement outcome \cite{PhysRevLett.104.073604}. This process can produce conditionally squeezed atomic momentum states that can be implemented in atom interferometers with significant metrological gain.

We quantify the attainable metrological gain $\xi_m$ by the ratio between the contrast squared $\mathcal{C}^2$ and the relative population variance $(\Delta N)^2$ normalized to the atom shot noise variance $N$ \cite{PhysRevA.50.67}:
\begin{equation}
\xi_m=\frac{N}{(\Delta N)^2}\mathcal{C}^2
\end{equation}
%-------------------------------------------------------
%        maybe this should be discussed later when the whole
%        system is well defined
%-------------------------------------------------------
%Free space scattering is the main limitation to the attainable squeezing. In our system, the main contribution derives from elastic Rayleigh scattering on the intercombination transition which spoils the coherence of the superposition state. As a result, it can be shown that the interferometer contrast decays as $\mathcal{C}=e^{-n_{\mathrm{sc}}}$, where $n_{\mathrm{sc}}$ is the number of free-space-scattered photons per atom \cite{PhysRevA.89.043837}.
 
The main limitation to the attainable squeezing is photon scattering into free space, which, in our system, is induced by the probe light itself and determines atom losses \cite{PhysRevA.79.023831}. After the scattering of one photon, an atom leaves the momentum superposition due to the recoil transferred in the photon scattering event. The result of these losses is a random imbalance $(\Delta N)_{\mathrm{sc}}$ of the populations in the two momentum states. Assuming that each atom scatters at most one photon, the population variance increase is $(\Delta N)_{\mathrm{sc}}^2=Nn_{\mathrm{sc}}$, where $N$ is the total atom number and $n_{\mathrm{sc}}<1$ is the number of free-space-scattered photons per atom.

As shown in Fig.~\ref{fig.figure1} b), in the presence of the probe field, atoms in the state $\ket{^1S_0,p}$ are coupled to the state $\ket{^1S_0,p+\hbar k_r}$, where $\hbar k_r$ is the probe photon momentum, corrected by the factor $\cos\alpha$ due to the angle between gravity and the oblique cavity beams (see Fig.~\ref{fig.figure1} a). When the probe light interacts with the momentum superposition, the atomic resonance is split in two lines at frequencies $\omega_1$ and $\omega_2$ corresponding to the transitions $\ket{^1S_0,0}-\ket{^3P_1,\hbar k_r}$ and $\ket{^1S_0,2n\hbar k_b}-\ket{^1S_0,2n\hbar k_b+\hbar k_r}$, respectively. In measuring momenta, here we assume that $p$ is measured in a reference frame moving at the same speed as the center of mass of the non-diffracted atomic cloud. The frequency splitting is $\omega_2-\omega_1\equiv 2\delta\omega_r =2n\hbar k_b k_r/M= 2\pi n\cos\alpha\times 28.6$ kHz, where $M$ is the atomic mass. For small $\alpha$, the factor $\cos\alpha$ yields a small correction to the frequency splitting which we neglect in our discussion. When the laser frequency $\omega_r$ is tuned halfway between the two optical transitions and on resonance with the cavity mode frequency $\omega_c$, atoms in the two momentum states produce phase shifts of opposite signs for the photons coupled into the cavity. 
In order to compute the photon phase shift $\Delta\phi_{\mathrm{ph}}$, we first consider a single optical transition at a frequency $\omega_0$ and write the expression for the cavity output field amplitude $\langle\hat{b}_{\mathrm{out}}\rangle$. In steady state, when the probe field interacts with $N$ atoms weakly driven through an input field with amplitude $\langle\hat{b}_{\mathrm{in}}\rangle = \beta$ \cite{PhysRevA.89.043837}, the output reflected field is
\begin{equation} \label{eq.transmitted_amplitude}
\braket{\hat{b}_{\mathrm{out}}}=\beta-i\frac{2\frac{\kappa_{\mathrm{in}}}{\kappa}\beta}{\frac{2\delta}{\kappa}+N\eta\mathcal{L}_d(\Delta)+i[1+N\eta\mathcal{L}_a(\Delta)]}.
\end{equation}
Here $\delta = \omega_r-\omega_c$ is the detuning of the probe lightù from the cavity resonance, $\Delta = \omega_r-\omega_0$ is the detuning from the atomic transition, $\kappa_{\mathrm{in}}$ is the photon loss rate from the input coupler, $\kappa$ is the total cavity photon loss rate, $\mathcal{L}_d(\Delta)=-2\Gamma\Delta/(\Gamma^2+4\Delta^2)$ and $\mathcal{L}_a(\Delta)=\Gamma^2/(\Gamma^2+4\Delta^2)$ are the atomic dispersion and absorption profiles, respectively, and $\Gamma$ is the atomic decay rate. The single-atom cooperativity is indicated as $\eta = 4g^2/(\Gamma\kappa)$, where $2g$ is the vacuum Rabi frequency.

\begin{figure}[tbp]\begin{center}
\includegraphics[width=0.47\textwidth]{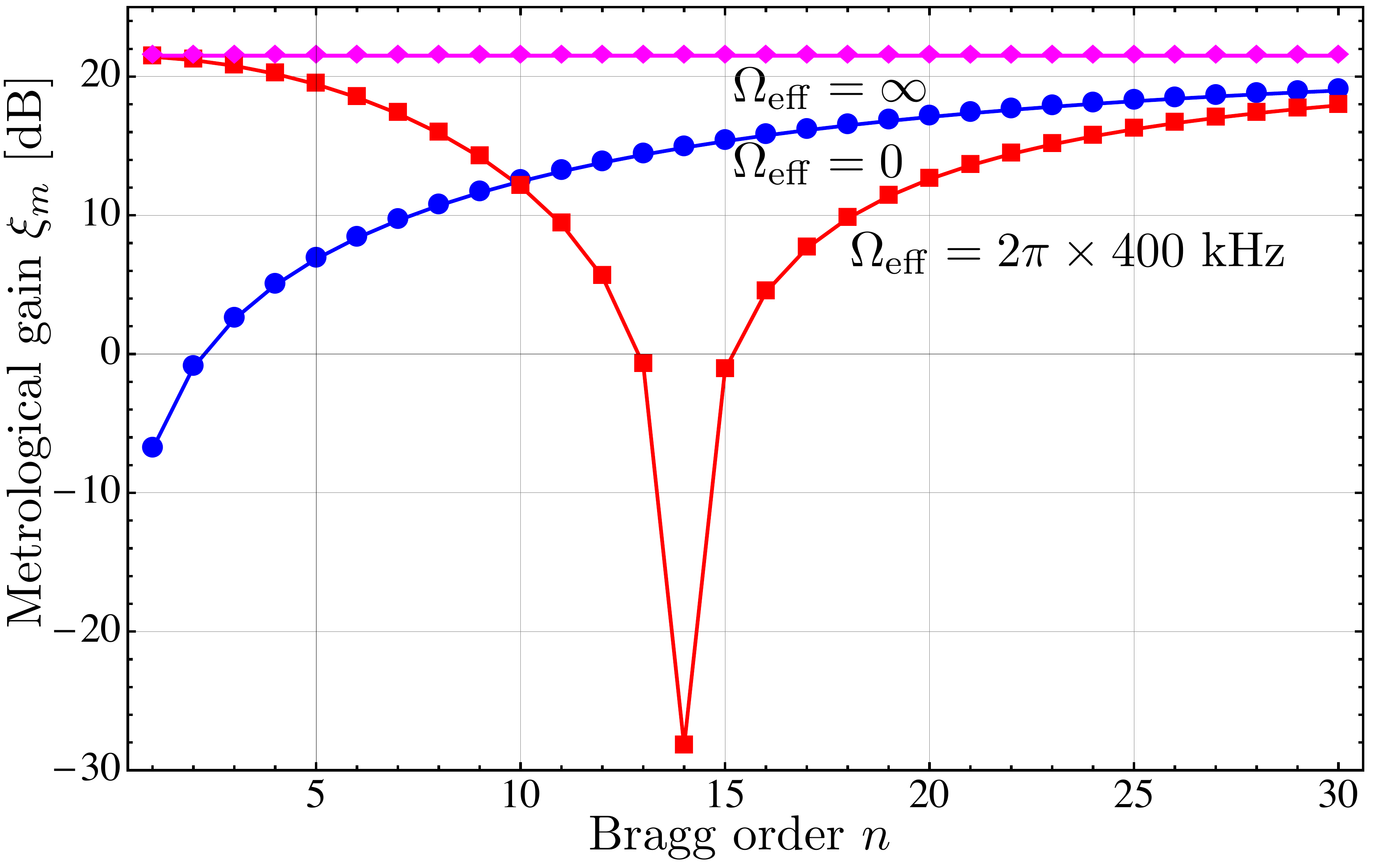}
\caption{Expected metrological gain $\xi_m$ as a function of the Bragg diffraction order $n$, assuming perfect detection efficiency, $\epsilon_d=1$ and for $N\eta = 10^4$. The gain is plotted when no EIT laser (see text) is present ($\Omega_{\mathrm{eff}}=0$, blue circles), when the EIT laser is present with finite Rabi frequency $\Omega_{\mathrm{eff}}=2\pi\times 400$ kHz (red squares) and in the ideal case of perfect transparency $\Omega_{\mathrm{eff}}=\infty$ (magenta diamonds).}
\label{fig.Bragg_squeezing}
\end{center}
\end{figure}

The phase shift $\Delta\phi_{\mathrm{ph}}$ can be computed through Eq. (\ref{eq.transmitted_amplitude}), by accounting for the presence of the two optical transitions and the initial state by the replacements $N\eta\mathcal{L}_{a,d}(\Delta)\rightarrow (N/2+\Delta N)\eta\mathcal{L}_{a,d}(\omega_r-\omega_1)+(N/2-\Delta N)\eta\mathcal{L}_{a,d}(\omega_r-\omega_2)$, which are equivalent to adding the corresponding polarizabilities and where $\Delta N$ is the population difference between the two momentum states.
The result is $\Delta\phi_{\mathrm{ph}} = 2\Delta N \eta\mathcal{L}_d(\delta\omega_r)/[1+N\eta\mathcal{L}_a(\delta\omega_r)]$, i.e. the population difference can be detected via the phase shift of the light emerging from the cavity. The phase shift measurement can be achieved through the setup illustrated in Fig.~\ref{fig.figure1} a), where the cavity-reflected light is measured. The phase measurement of the light is performed, for example, through the Pound-Drever-Hall technique. If the detector is performing at the photon shot noise level, the optimum atom number resolution, normalized to the variance of the atom shot noise, is given by
\begin{equation}
\frac{(\Delta N)^2}{N}=\frac{\mathcal{L}_a(\delta\omega_r)[1+N\eta\mathcal{L}_a(\delta\omega_r)]^2}{8N\eta\epsilon_d n_{\mathrm{sc}}[\mathcal{L}_d(\delta\omega_r)]^2},
\end{equation}
where $n_{\mathrm{sc}}$ is the number of photons scattered into free space per atom and $\epsilon_d$ is the detection efficiency.
By accounting for free space scattering we can then compute the optimum metrological gain, which is attained for
\begin{equation}
n_{\mathrm{sc}}=\sqrt{\frac{\mathcal{L}_a(\delta\omega_r)[1+N\eta\mathcal{L}_a(\delta\omega_r)]^2}{8N\eta\epsilon_d[\mathcal{L}_d(\delta\omega_r)]^2}}.
\end{equation} 
The resulting gain is plotted in Fig.~\ref{fig.Bragg_squeezing} for the case where $\epsilon_d=1$ and $N\eta = 10^4$, for varying Bragg diffraction orders.
When $N\eta$ lies in the range $10^3-10^4$ there is significant gain if $n>5$, a condition typically met by large-momentum-transfer atom interferometers \cite{Kovachy2015}. Indeed, for small $n$, the optical transitions are not sufficiently resolved in frequency space compared to the atomic linewidth, which prevents operating in the dispersive regime of atom-light interaction and leads to substantial absorption and squeezing reduction. In general, considerable gain can be observed if $N\eta\mathcal{L}_a(\delta\omega_r)\ll 1$, so that $n\gg 0.3\times\sqrt{N\eta}$, in which case the gain saturates at a value $\xi_m\approx \sqrt{2N\eta\epsilon_d}$, independent of the Bragg diffraction order.

%As illustrated in Fig.~\ref{fig.figure1} c), after producing the momentum superposition by a Bragg $\pi/2$ pulse and measuring the relative populations ($M_1$), the atomic state is projected into a state with reduced relative population uncertainty. The state on the Bloch sphere is then transformed into a phase-sensitive state by applying a Bragg $\pi/2$ pulse with a phase shift of 90$^{\circ}$ with respect to the first pulse. Following this preparation stage, a standard Mach-Zehnder interferometer sequence is applied. \textcolor{blue}{A final measurement ($M_2$) is performed using, for example, fluorescence detection.}

%-----------------------------------

%%\section{Resonant detection on a cycling transition}

%%\section{Conclusions}
%%We proposed and analysed two schemes that allow for the preparation of entangled squeezed states of motional degrees of freedom that can potentially be implemented as inputs in atom interferometers in a free falling configuration. The ring resonator design has interesting benefits compared to linear geometries, because the absence of standing waves for the probe field yields uniform entanglement and because the atomic cloud is free to exit the relatively small cavity volume and be injected in an atom interferometer. On the other hand, with strontium atoms, one can avoid systematic effects due to stray fields and observe long coherence times. The availability of narrow intercombination lines is a key feature that can allow to resolve pure momentum states by the Doppler effect. By limiting the detrimental effect of Raman scattering, metrological gain is limited by contrast loss due to elastic Rayleigh scattering and, with a measurement resolution at the single-atom level, one can attain Heisenberg-like scalings. With typical values for the single-atom cooperativity and atom number these schemes can induce more than 20 dB noise reduction compared to the atom shot noise limit.

\begin{figure}[tbp]\begin{center}
\includegraphics[width=0.47\textwidth]{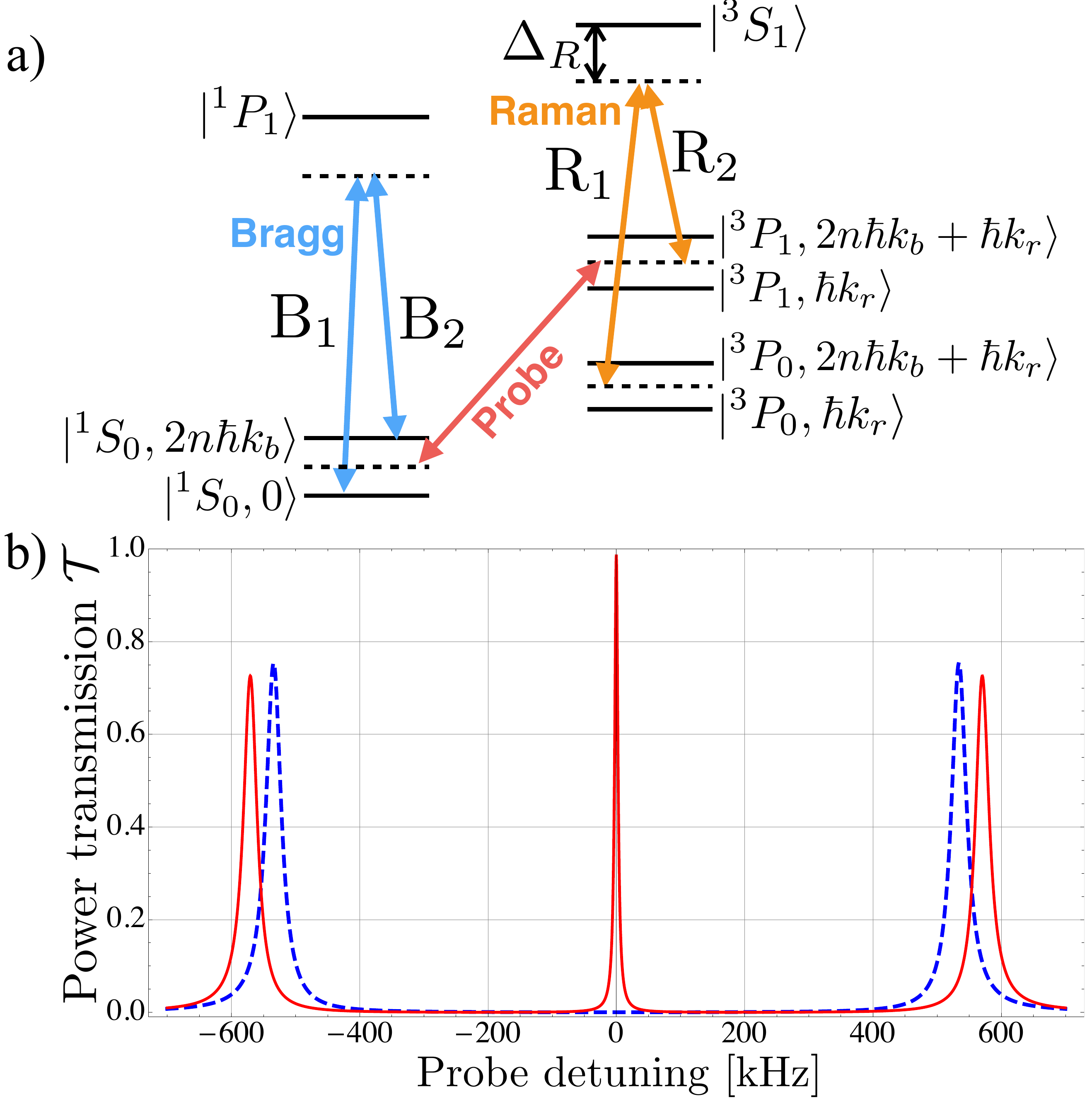}
\caption{a) Level diagram for the generation of squeezed momentum states enhanced by induced transparency via the $^3P_0$ state. The Bragg laser beams are indicated as $\mathrm{B}_1$ and $\mathrm{B}_2$. The effective control field that couples the states $^3P_0$ and $^3P_1$ is obtained by two-photon Raman coupling via the $^3S_1$ intermediate state. The two Raman beams $\mathrm{R}_1$ and $\mathrm{R}_2$ operate at the wavelengths 679 nm ($^3P_0-$$^3S_1$) and 688 nm ($^3P_1-$$^3S_1$), respectively, and are detuned from the transition to the $^3S_1$ state by $\Delta_R$. b) Cavity transmission spectrum with (red solid line) and without (blue dashed line) Raman coupling to the $^3P_0$ state for $\Omega_{\mathrm{eff}}=2\pi\times 400$ kHz, $N\eta=3\times 10^3$ and $n=1$. The population measurement is performed at the frequency of the transparency region (at probe detuning $\delta = 0$) which corresponds to a linewidth of $\kappa_{\mathrm{EIT}}=2\pi\times 6$ kHz.}
\label{fig.EIT_enhancement}
\end{center}
\end{figure}

We now show that it is possible to enhance the signal-to-noise ratio of momentum state population measurements also at small Bragg diffraction orders $n$, while keeping the collective cooperativity $N\eta$, and thus the squeezing, large. We therefore consider a scheme where we induce a transparency through electromagnetic fields (EIT) at the frequency of the probe laser \cite{PhysRevA.56.2385,PhysRevA.76.053814,PhysRevLett.100.173602,PhysRevLett.105.153603}.  To this end, we consider the coupling of the $^3P_1$ to the $^3P_0$ state as in Fig.~\ref{fig.EIT_enhancement} a). This is attained through two-photon Raman coupling via the $^3S_1$ intermediate state with the two copropagating Raman lasers $\mathrm{R}_1$ and $\mathrm{R}_2$ at 679 nm ($^3P_0-$$^3S_1$ transition) and 688 nm ($^3P_1-$$^3S_1$ transition), respectively. We define $\Omega_{\mathrm{R1}}$ and $\Omega_{\mathrm{R2}}$ as the single-photon Rabi frequencies of the two Raman beams and $\Delta_R$ as the detuning from the excited $^3S_1$ state. The effective control field is given by the two-photon Rabi frequency as $\Omega_{\mathrm{eff}}=\Omega_{\mathrm{R1}}\Omega_{\mathrm{R2}}/(2\Delta_R)$. We then find that the absorption and dispersion properties of the atomic medium are modified in such a way that the functions $\mathcal{L}_a(\Delta)$ and $\mathcal{L}_d(\Delta)$ are computed at the effective detuning $\Delta_E=\Delta-\Omega_{\mathrm{eff}}^2/(4\Delta)$, where we assumed that the Raman beams are on resonance with the two-photon transition $^3P_0-$$^3P_1$. 
In Fig.~\ref{fig.EIT_enhancement} b) we plot the spectrum of the cavity power transmission coefficient $\mathcal{T}$ for $N\eta=3\times 10^3$, $\Omega_{\mathrm{eff}}=2\pi\times 400$ kHz and diffraction order $n=1$. This shows that the coupling to the $^3P_0$ state results in an enhanced transmission at the frequency of the probe laser, allowing for a significant increase in the signal-to-noise ratio of the detected photons. 
EIT thus results in a reduced effective linewidth which in turn allows to operate in the dispersive regime. This condition is fulfilled when $N\eta\mathcal{L}_a(\delta\omega_E)\ll 1$, where now $\delta\omega_E=\delta\omega_r-\Omega_{\mathrm{eff}}^2/(4\delta\omega_r)$. We meet this condition when $|\delta\omega_E|/\Gamma\gg \sqrt{N\eta}/2$. We also note that in terms of laser power of the Raman beams, this condition is less demanding for narrow transitions compared to broad dipole-allowed transitions. 
The corresponding metrological gain in the presence of EIT coupling is shown in Fig.~\ref{fig.Bragg_squeezing} by the two lines corresponding to a finite coupling strength $\Omega_{\mathrm{eff}}=2\pi\times 400$ kHz and to the ideal infinitely strong coupling $\Omega_{\mathrm{eff}}=\infty$.

\begin{figure}[tbp]\begin{center}
\includegraphics[width=0.47\textwidth]{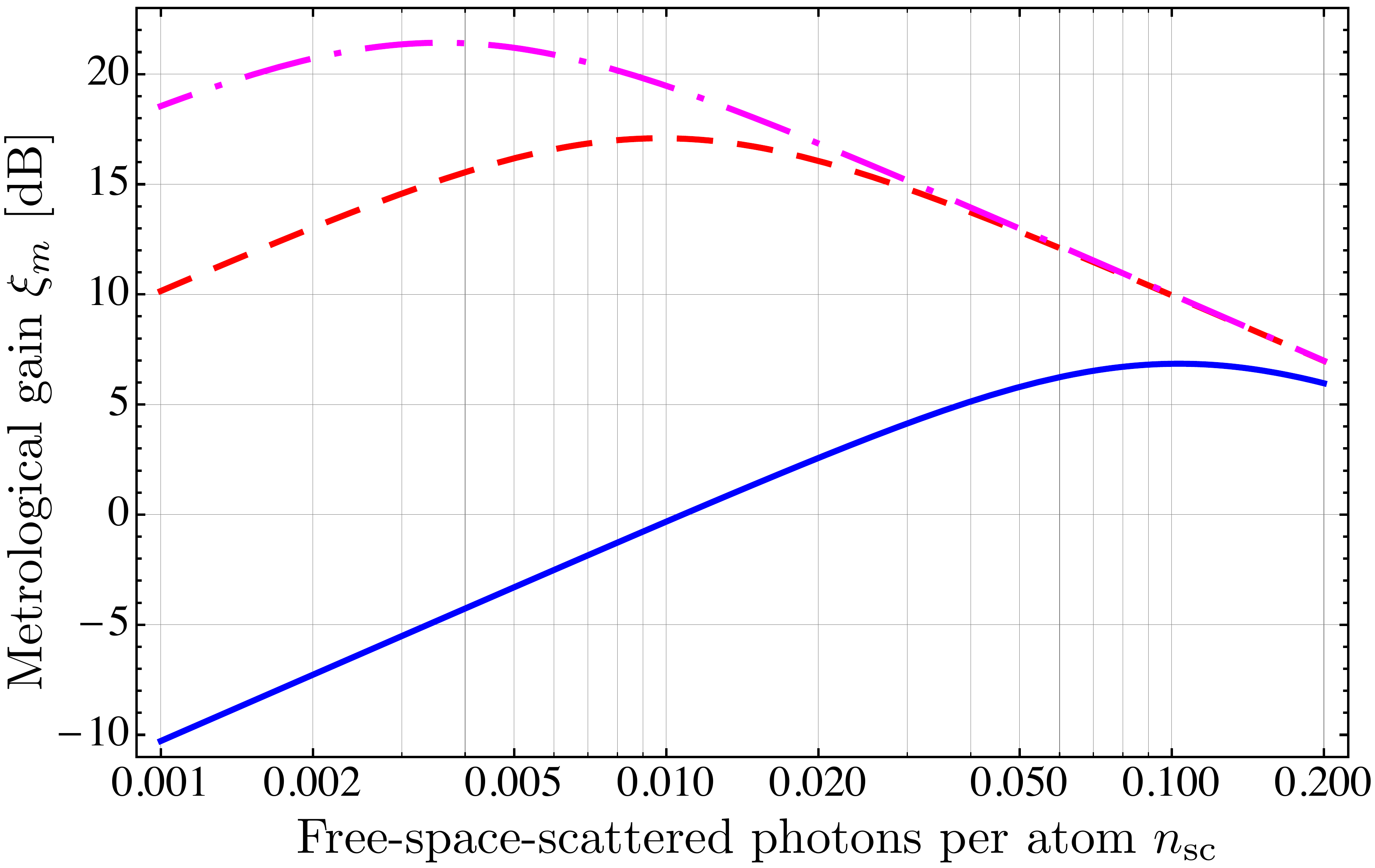}
\caption{Metrological gain $\xi_m$ as a function of the number of photons scattered into free space per atom for different Bragg diffraction orders and for varying Raman coupling strength: $n=5,\Omega_{\mathrm{eff}}=0$ (blue solid line); $n=20,\Omega_{\mathrm{eff}}=0$ (red dashed line); $n=1,\Omega_{\mathrm{eff}}=2\pi\times 400$ kHz (magenta dot-dashed line). Here we assume the collective cooperativity $N\eta=10^4$ and perfect detection efficiency $\epsilon_d = 1$.}
\label{fig.fig4}
\end{center}
\end{figure}

%\textcolor{red}{\textbf{See caption} In Fig.~\ref{fig.EIT_enhancement2} we plot the metrological gain as a function of the control field Rabi frequency $\Omega_{\mathrm{eff}}$ for different Bragg diffraction orders and for $N\eta=10^4$. We see that for $n=1$ and $\Omega_{\mathrm{eff}}> 2\pi\times 200$ kHz, the gain saturates at its optimum value, hence simulating the large $n$ case.}

In Fig.~\ref{fig.fig4} we summarize the difference between the presence and the absence of the Raman coupling to the $^3P_0$ state. This shows that, in terms of metrological gain, the EIT coupling is effectively equivalent to large diffraction orders and recovers the signal-to-noise ratio that would be otherwise lost because of photon absorption.

To summarize, the complete interferometric sequence, shown in Fig.~\ref{fig.figure1} c), is the following: strontium atoms are cooled and trapped in the waist close to the center of the optical cavity, then, a momentum superposition is created by
a Bragg $\pi/2$ pulse and immediately after that a measurement of the relative population is performed ($M_1$). At this stage the atomic ensemble is projected into a state with reduced relative population uncertainty.
The state on the Bloch sphere is then transformed into a phase-sensitive state by applying a Bragg $\pi/2$ pulse with a phase shift of 90$^{\circ}$ with respect to the first pulse.
Following this preparation stage, a standard Mach-Zehnder interferometer sequence is applied. A final measurement ($M_2$) is performed using, for example, fluorescence detection.

With realistic values for the various parameters, our method is applicable to strontium atoms with the current technology. Specifically, we consider an optical cavity where one of the foci has a waist $w_0=150\mbox{ }\mu$m, at the position where the atoms cross the cavity mode volume. With a cavity finesse $F=2.5\times 10^4$ and at the wavelength $\lambda=689$ nm, we get a single-atom cooperativity $\eta = 6F\lambda^2/(\pi^3w_0^2)\approx 0.1$. We then consider $N\approx 10^5$ atoms occupying a volume with a size of about $30\mbox{ }\mu$m. With these parameters, the collective cooperativity is $N\eta\approx 10^4$. 
The maximum Bragg diffraction order allowable for our method is set by the condition that the transit time of the two clouds corresponding to the momentum states through the cavity beam waist is larger than the time duration of the collective measurement. We estimate the useful transit time as the one taken by a cloud with speed $n\hbar k/M$ to cross one tenth of the effective mode waist. Because the atoms are crossing the cavity beam vertically, the effective mode waist is $w_0/\sin\alpha$. We therefore estimate the maximum Bragg diffraction order as $n_{\mathrm{max}}=Mw_0/(10\hbar k T_m\sin\alpha)$, where $T_m$ is the measurement time duration. With $\alpha\approx 0.4$ rad and $T_m\approx 200\mbox{ }\mu$s we get $n_{\mathrm{max}}=10$. However, the maximum Bragg order can be made considerably larger by a suitable design of the cavity geometry, where $w_0$ is made larger and $\alpha$ is made smaller.
The measurement time is set by the requirement that the number of photons scattered into free space is sufficient to provide the optimum metrological gain. 
%\textcolor{red}{We estimate the rate of photons scattered into free space per atom as $\dot{n}_{\mathrm{sc}}= \Gamma P_{\mathrm{exc}}$, where  $P_{\mathrm{exc}}$ is the population of the excited $^3P_1$ state. By setting $P_{\mathrm{exc}}= 5\times 10^{-4}$ and $T_m=200\mbox{ }\mu$s, we achieve the optimum gain at $n_{\mathrm{sc}}\approx 5\times 10^{-3}$. By assuming a finite detection efficiency $\eta_d=1/2$, first order Bragg diffraction ($n=1$) and a Raman Rabi frequency $\Omega_{\mathrm{eff}}=2\pi\times 400$ kHz, we conclude that it is possible to achieve about 20 dB of metrological gain.}

By considering the collective cooperativity $N\eta = 10^4$, the Raman coupling strength $\Omega_{\mathrm{eff}}=2\pi\times 400$ kHz, a measurement time $T_m=200\mbox{ }\mu$s and a finite detection efficiency $\epsilon_d=1/2$, we conclude that the optimum number of photons scattered into free space per atom is $n_{\mathrm{sc}}=5\times 10^{-3}$, corresponding to the excited state population $P_{\mathrm{exc}}=n_{\mathrm{sc}}/(\Gamma T_m)=5\times 10^{-4}$. In this case it is possible to achieve a metrological gain of 20 dB.

In conclusion, we have proposed and analysed a novel scheme that allows for the production of squeezed momentum states for large-momentum-transfer Bragg atom interferometers.
The essence of our method is based on the ability to resolve the Doppler splitting of two momentum states by using a probe laser with frequency close to the narrow $^1S_0-$$^3P_1$ resonance of strontium.
With realistic parameters we show that about 20 dB of noise reduction in atom interferometer phase measurements can be attained compared to the Standard Quantum Limit, after less than 1 ms for the preparation stage. 

Moreover, at small Bragg orders, where cavity-enhanced absorption spoils the resolution necessary for the measurement, we have shown that it is possible to implement two-photon Raman coupling in order to induce a transparency at the frequency of the probe laser, therefore recovering the dispersive regime of atom-light interaction. 

With this method it is possible to attain significant squeezing also for small Bragg diffraction orders and large atom numbers, with a scaling for the interferometer phase resolution $\Delta\phi\sim N^{-3/4}$, where $N$ is the atom number.
Our method is applicable to other atomic species with narrow transitions.

\section*{Acknowledgements}
Maria Luisa Chiofalo and Alice Sinatra are acknowledged for useful discussions. This work was supported by INFN and the Italian Ministry of Education, University and Research (MIUR) under the Progetto Premiale ``Interferometro Atomico'' and the PRIN 2015 project ``Interferometro Atomico Avanzato per Esperimenti su Gravit\`{a} e Fisica Quantistica e Applicazioni alla Geofisica''. V.V. acknowledges support by NSF, NSF CUA, and ONR.

%\bibliography{bib}{}


\end{document}